\documentclass{aa}  
\PassOptionsToPackage{comma}{natbib}
\usepackage{natbib}
\makeatletter
\renewcommand\NAT@sep{, }
\makeatother
\usepackage{microtype}
\usepackage{amssymb,amsmath}
\usepackage{graphicx}
\usepackage{multirow}
\usepackage[T1]{fontenc}
\usepackage{ae,aecompl}
\usepackage{newtxtext,newtxmath}
\usepackage{longtable,listings}
\usepackage{natbib,twoopt}
\usepackage{graphicx}
\usepackage{soul, color, xcolor}
\usepackage{hyperref}

\titlerunning{gravitational acceleration model}
\authorrunning{Chen et al.}

\begin{document} 
\title{A gravitational acceleration model to explain the double-peaked narrow emission lines shifted in the same direction}
\author{XingQian Chen\inst{1}\and GuiLin Liao\inst{1}\and Qi Zheng\inst{1} \and XueGuang Zhang \inst{1}\fnmsep\thanks{E-mail: {xgzhang@gxu.edu.cn}}}
\institute{School of Physical Science and Technology, Guangxi University, No. 100, Daxue East Road, Nanning 530004, P. R. China}
\date{}

  \abstract
  {In this manuscript, we propose, for the first time, an oversimplified but potentially effective gravitational acceleration model to interpret 
the double-peaked narrow emission lines (DPNELs) shifted in the same direction. 
We adopt the framework of a merging kpc-scale dual-core system in an elliptical orbit, which has 
an emission-line galaxy with clear narrow line regions (NLRs) merging with a companion galaxy lacking emission line features. 
Due to gravitational forces induced by both galaxies on the NLRs, the accelerations of the far-side and near-side NLR components may share the 
same vector direction when projected along the line-of-sight, leading the velocities of the observed DPNELs to shift in the same direction. 
Our simulations indicate that the probability of producing double-peaked features shifted in the same direction reaches 5.81\% in merging 
kpc-scale dual core systems containing emission-line galaxies. 
Besides the expected results from our proposed model, we identify a unique galaxy SDSS J001050.52-103246.6, 
whose apparent DPNELs shifted in the same direction can be plausibly explained by the gravitational acceleration model. 
This proposed model provides a new path to explain DPNELs shifted in the same direction in the scenario that the two galaxies 
align along the line-of-sight in kpc-scale dual-core systems.
}


   \maketitle
%

\section{Introduction}

Double-peaked narrow emission lines (DPNELs) have been investigated since the 1980s, especially in active galactic nuclei (AGN), 
as discussed in \cite{1981ApJ...247..403H, 1984ApJ...281..525H, 1985Natur.318...43K}, and \cite{1991ApJS...75..357V}.
These features in the spectra are commonly attributed to two kinematically distinct emission components with different velocities, 
as discussed in \cite{2007ApJ...660L..23G, 2009ApJ...705L..76W}, and \cite{2010ApJ...708..427L}. 
Since DPNELs galaxies play a critical role in studying the dynamics of AGN narrow line regions (NLRs) and merging galaxies, 
they have motivated the establishment of large and homogeneous samples to investigate their properties through statistical 
analyses given their unclear physical origins (\cite{2010ApJ...716..866S, 2012ApJS..201...31G, 2012ApJ...757..124W, 2015ApJ...813..103M, 
2018ApJ...867...66C, 2019MNRAS.482.1889W}, and \cite{2025ApJS..277...49Z}).  

At the current stage, the physical origins of DPNELs can be categorized into three primary scenarios. 
Firstly, DPNELs can arise from biconical geometry and disturbed gas kinematics in the NLR, driven by AGN radiation pressure, as discussed in \cite{2011ApJ...727...71F, 2011ApJ...735...48S, 2012ApJ...745...67F}, and \cite{2013ApJ...769...95B}. 
Secondly, rotating disk-like NLR structures in AGN can produce DPNELs due to the orbital motion of ionized gas, 
as proposed in \cite{2011ApJ...739...69M, 2015ApJ...799..161K}, and \cite{2023A&A...670A..46M}.
Thirdly, DPNELs can stem from the orbital motion of kiloparsec scales (kpc-scale) dual-core systems containing independent NLRs, as discussed in 
\cite{2004ApJ...604L..33Z}, \cite{2009ApJ...705L..20X}, and \cite{2011ApJ...740L..44F}.

In conventional DPNELs, the two emission components arising from either NLRs kinematics or dual-core systems exhibit similar shifted emission 
line features, one of the peaks is redshifted and the other peak is blueshifted relative to the redshift of the host galaxy.
However, a peculiar subclass of DPNELs displays both emission components shifted in the same direction, a 
phenomenon unexplained by existing scenarios. 
We propose a gravitational acceleration model to elucidate its physical origin for the first time. 
However, this subclass of DPNELs may arise from inaccurate determinations of the host galaxy redshift, 
indicating the need for reliable redshift measurements.

In Section 2, we present our main hypotheses of the gravitational acceleration model. In Section 3, we perform our main simulations, and discussions on 
SDSS J001050.52-103246.6 with the DPNELs shifted in the same direction. 
Main summary and conclusion are presented in Section 4. 
Throughout this manuscript, we adopt the cosmological parameters \( H_0 = 70 \, \text{km s}^{-1} \text{Mpc}^{-1} \), \( \Omega_M = 0.3 \), 
and \( \Omega_\Lambda = 0.7 \).

\section{MAIN HYPOTHESES}
\subsection{gravitational acceleration model}
\begin{figure}
 \includegraphics[width=0.90\columnwidth]{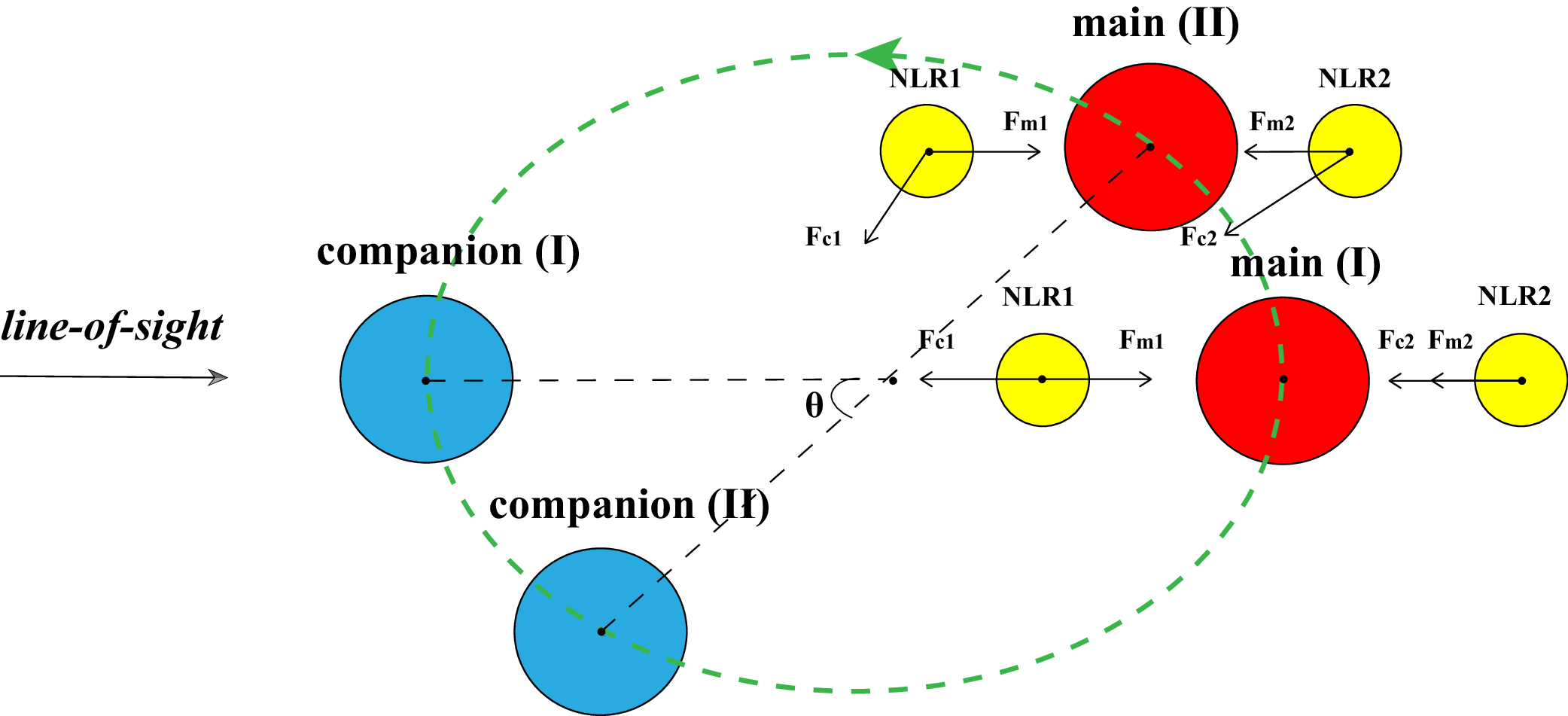}
 \centering
 \caption{The diagram of the gravitational acceleration model with the two galaxies initially located at the apocenter of the 
elliptical orbit.
The circles filled in red and blue represent the main and companion galaxies, respectively. 
The circles filled in yellow represent the NLRs of the main galaxy. The black arrows represent the gravitational force vectors. 
The green dashed line represents the clockwise elliptical orbit, with $\theta$ marking the rotation angle. 
System I indicates the initial positions of the galaxies in the simulation, aligned along the line-of-sight, 
while System II indicates their final positions. 
The arrow in the left region marks the line-of-sight. 
}
 \label{model_figure}
\end{figure}

The firstly proposed gravitational acceleration model is based on a kpc-scale dual-core system, where the two galaxies orbit each other in 
an elliptical trajectory at kpc-scale separations. 
The emission-line galaxy containing the far-side and near-side NLR components is designated as the main galaxy, while the orbiting 
non-emission-line galaxy is designated as the companion galaxy. 
We assume that the two galaxies are initially located at the apocenter of the elliptical orbit and are aligned along the line-of-sight.
Adopting the ionized cone model of NLRs from \cite{2013MNRAS.430.2327L}, we assume the NLRs are symmetric mass distributions positioned on 
opposite sides of the main galaxy. 
Figure \ref{model_figure} illustrates the model configuration.

Based on the law of universal gravitation and Newton's laws of motion, the near-side (NLR1) and far-side (NLR2) of NLRs, 
under the gravitational forces of both main and companion galaxies, yield individual accelerations, which can be described as:
\begin{equation}
a_{1/2} = \frac{GM_{c}}{d_{c1/c2}^2} \pm \frac{GM_{m}}{d_{m1/m2}^2}, 
\end{equation}
where $M_{c}$ and $M_{m}$ are the total stellar masses of the main and companion galaxies; $d_{c1}$, $d_{c2}$, $d_{m1}$ and $d_{m2}$ denote 
the distances from the companion/main galaxy to NLR1/NLR2. 
The gravitational forces between NLRs are not considered in the acceleration computations, nor is there any consideration of the 
gas mass of the companion galaxy in estimating its total mass. Detailed discussions can be found in Appendix \ref{NLR_mass} and \ref{gas_mass}.
Here, the leftward direction is defined as positive. 
Both accelerations can thus attain positive values (indicating the same vector direction) with different magnitudes after projecting along the line-of-sight. 
Consequently, a period of acceleration (hereafter acceleration duration) causes both NLRs to obtain velocities shifted in the same direction. 
The acceleration process always requires that the distances between NLRs and galaxies exceed 0.05 kpc.

\begin{figure*}
 \includegraphics[width=0.80\paperwidth]{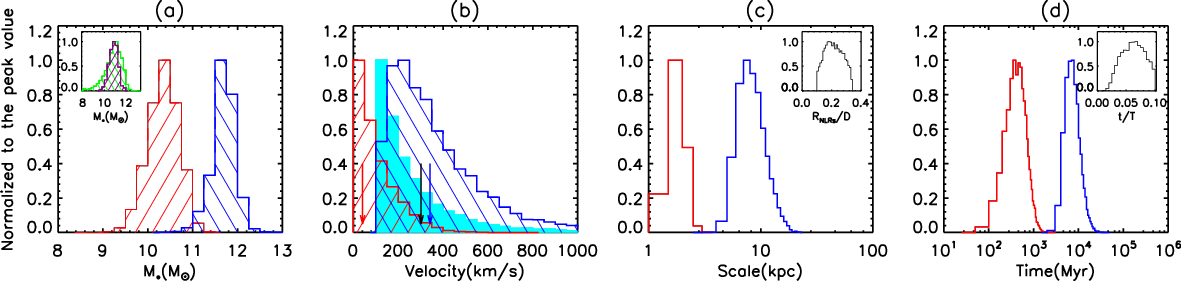}
 \centering
 \caption{Simulation results for the model adopting an elliptical orbit with an eccentricity of 0.2 and with galaxies initially at apocenter.  
Panel (a): the histograms filled in red and blue show the logarithmic total stellar mass distributions of the main and companion 
galaxies in the 10,000 satisfied simulations with DPNELs shifted in the same direction. 
Here, the top left inset displays the mass distributions for the galaxies in the collected main-sample (histogram in purple) and companion-sample 
(histogram in green).  
Panel (b): the histograms filled in red, blue, and cyan show the distributions of the smaller blueshifted velocity, larger 
blueshifted velocity, and velocity separation of DPNELs shifted in the same direction; 
the arrows in red, blue, and black mark the smaller blueshifted velocity, larger blueshifted velocity, and velocity separation of SDSS J001050.52-103246.6. 
Panel (c): the histograms in red and blue show the distributions of $R_{NLRs}$ and galaxy separation; the distribution 
of $R_{NLRs}$/D is shown in the top right region. 
Panel (d): the histograms filled in red and blue show the distributions of the acceleration duration and orbital period; the distribution 
of the acceleration duration/orbital period is shown in the top right region. 
All the histograms are normalized to the peak value. 
}
 \label{total_dis}
\end{figure*}

\begin{figure}
 \includegraphics[width=0.85\columnwidth]{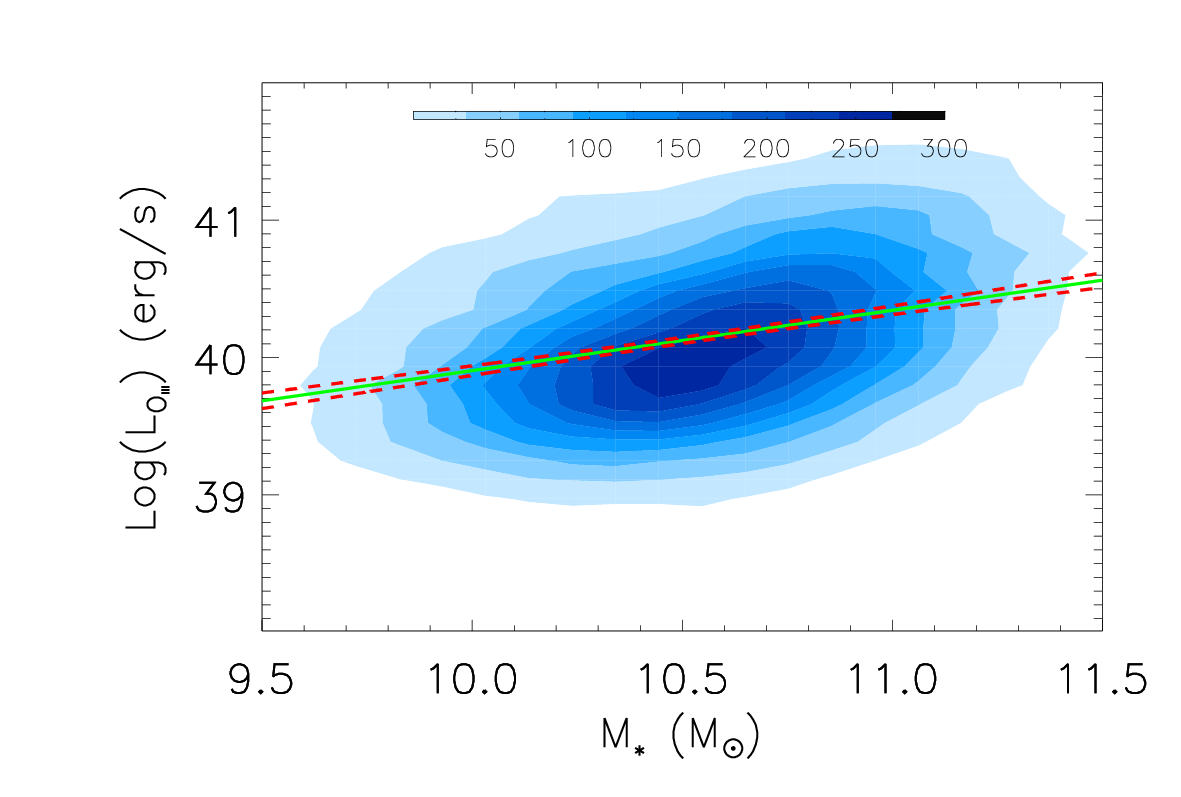}
 \centering
 \caption{On the correlation between $L_{O~\textsc{iii}}$ and logarithmic total stellar mass for the galaxies in the main-sample. 
The green solid line indicates the best fitting results, and the red dashed lines indicate the corresponding 5$\sigma$ confidence bands. 
Contour levels in different colours are related to number densities shown in colour bar in the top region.
}
 \label{mass_loiii}
\end{figure}

\subsection{simulation parameters}
To determine the parameters required for performing simulations based on the gravitational acceleration model, we adopt the following methods 
to derive reasonable parameters.

We utilize the Structured Query Language (SQL) search tool provided by Sloan Digital Sky Survey (SDSS) to collect galaxy samples for the
main (main-sample) and companion (companion-sample) galaxies to determine their total stellar masses. 

For the main-sample, the emission line fluxes of H$\alpha$, H$\beta$, [O~\textsc{iii}]$\lambda$5007\AA, [N~\textsc{ii}]$\lambda$6584\AA, 
and [S~\textsc{ii}]$\lambda$6717, 6731\AA\ doublet, along with corresponding uncertainties are selected from the SDSS $GalSpecLine$ database, 
and the total stellar masses (the parameter $mstellar$\_$median$ and corresponding uncertainty $mstellar$\_$err$) are selected from 
the $stellarMassPCAWiscM11$ database.  
The selection criteria for the main-sample are provided in Appendix \ref{main_selection}, yielding 159,911 galaxies in the main-sample. 

For the companion-sample, the selected parameters and the first three criteria are the same as those for the main-sample. 
An additional requirement imposes flux < 3*flux\_err for the same emission lines, ensuring negligible emission line features. 
This results in 8.916 galaxies being collected in the companion-sample.  
The distributions of the logarithmic total stellar masses of both samples are shown in the top-left region of Panel (a) in Figure \ref{total_dis}.

Given that the total stellar mass of the main galaxy can be determined from the main-sample, 
the size of the NLRs (the distance between NLRs and central black hole, hereafter $R_{NLRs}$) should also be reasonably determined. 
Therefore, we combine the known empirical $R_{NLRs}-L_{[O~\textsc{iii}]}$ correlation in AGNs from \cite{2013MNRAS.430.2327L} 
($L_{[O~\textsc{iii}]}$ as the [O~\textsc{iii}] 5007\AA\ luminosity) with the $L_{[O~\textsc{iii}]}-M_{*}$ correlation derived from our 
main-sample ($M_{*}$ as the logarithmic total stellar mass), to determine $R_{NLRs}$ via total stellar mass. 
The $R_{NLRs}-L_{[O~\textsc{iiii}]}$ correlation in the literature is expressed as:
\begin{equation}
log(\frac{R_{NLRs}}{pc}) = (0.25 \pm 0.02) \times log(\frac{L_{[O~\textsc{iii}]}}{10^{42}erg/s}) + (3.75 \pm 0.03). 
\end{equation}
Using the collected main-sample, we can estimate the $L_{[O~\textsc{iii}]}-M_{*}$ correlation: 
\begin{equation}
log(L_{O~\textsc{iii}}/erg/s) = \alpha \times M_{*} + \beta, 
\end{equation}
through the Least Trimmed Squares regression technique, the best-fitting parameters are measured: $\alpha$=0.44$\pm$0.01 
and $\beta$=35.52$\pm$0.06 (the 1$\sigma$ uncertainties are determined through this technique), with a rank correlation coefficient of 0.34. 
Figure \ref{mass_loiii} plots the best-fitting results. 
Although our main-sample comprises emission-line galaxies rather than AGNs, 
we retain this approach to minimize free parameters in simulations. 
Therefore, with the help of the two correlations, an accepted dependence of $R_{NLRs}$ on total stellar mass is established. 

The separation between the main and companion galaxies (hereafter galaxy separation) is assumed to be 3$\sim$10 times $R_{NLRs}$, 
leading to a maximum value of $\sim$35 kpc. 
Detailed descriptions of the determination of galaxy separation are provided in Appendix \ref{galaxy_separation}.
To neglect the orbit decay, the acceleration duration is constrained to be less than 10\% of the orbital period of the kpc-scale 
dual-core system, limiting the maximum rotation angle to $\theta$ < 36\textdegree. 
The orbital period of the elliptical orbital motion is calculated via Kepler's third law:
\begin{equation}
\frac{r^{3}}{T^{2}} = \frac{G(M_{m}+M_{c})}{4\pi^2}, 
\end{equation}
where $r$ is the semi-major axis, measured as D/(1+e) (D as the galaxy separation and e as the eccentricity of the elliptical orbit); 
$M_{m}$ and $M_{c}$ are the total stellar masses of the main and companion galaxies, respectively; $T$ is the orbital period. 
Additionally, we impose a minimum projected velocity separation of 100 km/s between DPNELs peaks to ensure resolvable 
double-peaked profiles. Detailed descriptions are provided in Appendix \ref{velocity_separation}.

Moreover, to possibly recreate the realistic galaxy merging process, 
we enforce two NLR displacement constraints to prevent mergers between galaxies and NLRs. 
The distance between the far-side NLR (NLR2) and the center of the main galaxy is at least 0.05 kpc ($d_{m2}$ > 0.05 kpc), 
and the near-side NLR (NLR1) is confined in half the galaxy separation ($d_{m1}$ < D/2). 

\section{MAIN RESULTS}
\subsection{simulation results}
Building on the hypotheses outlined in Section 2, we introduce the workflow and key parameters of our oversimplified simulations, 
based on the gravitational acceleration model adopting an elliptical orbit with an eccentricity of 0.2 (as an example) and with 
the two galaxies initially located at the apocenter of the elliptical orbital path. 
In each simulation, the total stellar masses of the main and companion galaxies are randomly drawn from the main-sample and 
companion-sample, respectively. 
The $R_{NLRs}$ of the main galaxy is then calculated via the $R_{NLRs}-L_{[O~\textsc{iiii}]}$ and $L_{[O~\textsc{iii}]}-M_{*}$ correlations 
(Section 2.2). 
The corresponding uncertainties of both correlations are also incorporated in the computational process, 
the effects of adopting the uncertainties on the simulation results are discussed in Appendix \ref{correlation_uncertainty}.
The galaxy separation is determined by multiplying $R_{NLRs}$ by a random factor between 3 and 10. 
The acceleration duration is randomly assigned between 0\% and 10\% of the orbital period, which is computed using Equation (4) with the 
selected total stellar masses and the semi-major axis determined by the galaxy separation.
During the acceleration phase, the duration is divided into 50 time segments, with the shifted velocities and spatial extensions of 
the NLRs calculated in each segment. 
We retain simulations where the velocity separation exceeds 100 km/s and NLR extensions satisfy the NLR displacement constraints.

Through 172,000 simulation trials, 10,000 runs satisfy our selection criteria, indicating that approximately 5.81\% of kpc-scale dual-core systems with 
emission-line galaxies may exhibit DPNELs shifted in the same direction. 
Figure \ref{total_dis} presents the simulation results related to the galaxies with DPNELs shifted in the same direction. 
The logarithmic total stellar mass distributions of the main and companion galaxies differ significantly, with mean 
values of 10.35 and 11.69, respectively. 
The velocity separation has a mean value of 321 km/s, slightly higher than the mean velocity separation of common DPNELs 
of 300 km/s reported in \cite{2012ApJS..201...31G}. 
The $R_{NLRs}$ span from 1.04 kpc to 3.12 kpc, while galaxy separations extend to 3.75$\sim$22.07 kpc. 
The acceleration duration varies from 0.53 Myr to 22.53 Myr, while the orbital period is in the range of 22.43$\sim$462.12 Myr.

\begin{figure*}
 \includegraphics[width=0.7\paperwidth]{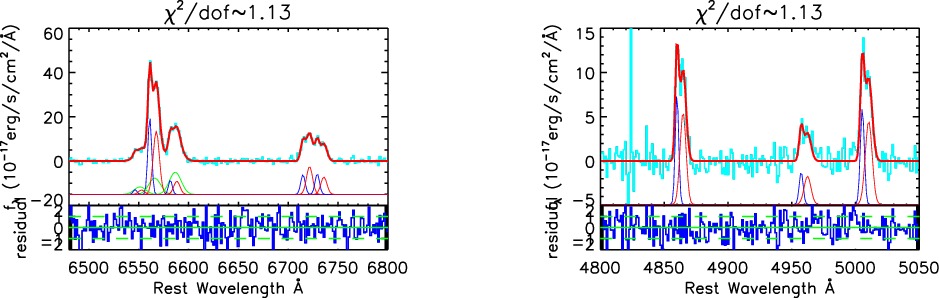}
 \centering
 \caption{Spectroscopic results of SDSS J001050.52-103246.6. 
The left and right panels present the best-fitting results to the emission lines around H$\alpha$ and H$\beta$, respectively.
In both panels, the cyan components indicate the spectrum after subtracting the host galaxy starlight; the thick 
red components indicate the best-fitting results; the solid lines in the bottom region of each panel indicate residuals. 
In the left panel, the thin red and thin blue Gaussian components from left to right indicate the double-peaked features of narrow  
[N~\textsc{ii}]$\lambda$ 6550\AA, H$\alpha$, [N~\textsc{ii}]$\lambda$ 6585\AA, [S~\textsc{ii}]$\lambda$ 6718\AA, and [S~\textsc{ii}]$\lambda$ 6732\AA; 
the green Gaussian components indicate the extended emission components in [N~\textsc{ii}]$\lambda$ 6550\AA, H$\alpha$, 
and [N~\textsc{ii}]$\lambda$ 6585\AA. 
In the right panel, the thin red and thin blue Gaussian components from left to right indicate the double-peaked features of narrow H$\beta$, 
[O~\textsc{iii}]$\lambda$ 4959\AA, and [O~\textsc{iii}]$\lambda$ 5007\AA. 
In each panel, the corresponding $\chi^2$/dof related to the best-fitting results is marked in the title. 
}
 \label{spectrum}
\end{figure*} 

\subsection{an instance with DPNELs shifted in the same direction}
To identify galaxies exhibiting DPNELs shifted in the same direction as a case study, we searched SDSS DR16 and identified one object
SDSS J001050.52-103246.6 (hereafter SDSS J0010).
We analyzed its spectroscopic data using methodologies similar to our prior work in \cite{2025MNRAS.539L..24C}. 
The detailed spectroscopic handling procedure of SDSS J0010 is provided in Appendix \ref{spectroscopic_analysis}.

The spectroscopic results of SDSS J0100 are shown in Figure \ref{spectrum}.
Consequently,  we obtain the velocity parameters of DPNELs of SDSS J0100,  with blueshifted velocities of -343.32$\pm$17.27 km/s and 
-41.51$\pm$36.71 km/s, and a velocity separation of $301.80^{+26.99}_{-9.72}$ km/s. 
These values are annotated in Panel (b) of Figure \ref{total_dis} with colored arrows. 
However, as its blueshifted velocity offsets and velocity separation fully align with our simulated parameter space, its DPNELs 
shifted in the same direction can be interpreted by our model from a purely numerical simulation perspective.
Detailed discussions of DPNELs shifted in the same direction of SDSS J0010 are provided in Appendix \ref{spectroscopic_analysis}.
Further, a sample consisting of galaxies similar to SDSS J0010 with such peculiar DPNELs features is prepared by us.

\subsection{discussions of stated assumptions}
Figure \ref{model_figure} depicts the projection-aligned scenario. 
In reality, inclination exists between the galaxy pair and between the galaxies and NLRs. 
Considering that only gravity and velocity components perpendicular to the line-of-sight contribute to the final results, 
we assume the NLRs and galaxies are initially aligned along the line-of-sight.
Additionally, the discussions of the model configuration reality are provided in Appendix \ref{model_reality}.

Galaxy mergers inherently involve orbital velocities. 
In our framework, the NLRs and the main galaxy share identical orbital velocities, resulting in equal contributions to the shifted  
velocities of the near-side and far-side NLRs. 
While this affects the magnitude of blueshifted velocities but not the velocity separation, the orbital velocity is not included in the 
computational procedure.

Our model assumes an elliptical orbit for the merging galaxy pair. 
The simulation results presented in Figure \ref{total_dis} are simulated by the model in an elliptical orbit with an eccentricity of 0.2 
and with galaxies initially at apocenter.
We also explore the effects of different eccentricities of the elliptical orbital path and the initial positions where the galaxies are 
located on the final simulation results.
The detailed discussions are provided in Appendix \ref{different_path}. 

If both galaxies have emission line features, their mutual gravitational interactions in a kpc-scale dual-core system 
could theoretically produce four-peaked narrow emission lines.
It will be an interesting challenge to detect the galaxies with four-peaked narrow emission lines in the future.

\section{CONCLUSION}
We first propose a brand new gravitational acceleration model to explain a special subclass of DPNELs exhibiting velocities shifted in the same direction. 
Through an oversimplified simulation of 172,000 trials, 10,000 cases meet the criteria for generating such features, implying a 5.81\% probability of 
arising DPNELs shifted in the same direction in merging kpc-scale dual core systems containing emission-line galaxies.
Meanwhile, our model plausibly explains the DPNELs shifted in the same direction of SDSS J001050.52-103246.6, as its
blueshifted velocity offsets and velocity separation fully align with our simulated parameter space. 
In the near future, it is one of our main objectives to detect and discuss a sample of such DPNELs collected from SDSS.

\section*{ACKNOWLEGEMENTS}
The authors gratefully acknowledge the anonymous referee for giving us constructive comments and suggestions to greatly 
improve the paper.
Zhang gratefully acknowledges the kind financial support from GuangXi University, and the grant support from NSFC-12173020 and 
NSFC-12373014.
Chen gratefully acknowledges the kind grant support from Innovation Project of Guangxi Graduate Education YCSW2024006.
This manuscript has made use of the data from SDSS projects. The SDSS-III website is \url{http://www.sdss3.org/}. 
The SDSS DR16 website is \url{http://skyserver.sdss.org/dr16/en/home.aspx}.

\appendix
\section{Extra explanations}

\subsection{why we do not consider the gravitational forces between NLRs}\label{NLR_mass}
In calculating the gravitational accelerations of the two NLRs of the main galaxy, we mainly consider the gravitational forces 
generated by the two galaxies.
According to the classical textbook of \cite{1989agna.book.....O}, the NLR masses are approximately $10^{5\sim6} M_\odot$, which 
are negligible compared to the galaxy masses (nearly $10^8 \sim 10^{13} M_\odot$) in our galaxy samples. 
Thus, the gravitational forces between NLRs are substantially weaker than those induced by the galaxies, the forces between NLRs are 
therefore excluded in the acceleration computations.

\subsection{why the gas mass of the companion galaxy is not considered}\label{gas_mass}
In our model, the gas of the companion galaxy is not considered in estimating the total mass of the companion galaxy.
According to \cite{2018ApJ...864...40P}, the correlation between stellar mass and H~\textsc{i} mass is given by:
\begin{equation}
logM_{H~\textsc{i}} = 0.51(logM_{*} - 10) + 9.71,
\end{equation}
where $log M_{H~\textsc{i}}$ represents logarithmic H~\textsc{i} mass and $logM_{*}$ represents logarithmic stellar mass. 
Based on the assumption that the gas mass equals H~\textsc{i} mass, 
we performed simulations incorporating the gas mass of the companion galaxy, determined by its stellar mass. 
The results show a mean velocity separation of $\sim$325 km/s, slightly higher than the value (321 km/s) from simulations neglecting 
the gas mass of the companion galaxy. 

\subsection{selection criteria for the main-sample}\label{main_selection}
The following selection criteria are applied to collect a reasonable sample for the main galaxy:
i) class='galaxy', to guarantee the SDSS spectral classification;
ii) z < 0.3 and z.warning = 0, to ensure spectral coverage of the [S~\textsc{ii}] doublet and reliable redshift determination; 
iii) $mstellar$\_$err$ > 0 and $mstellar$\_$median$ > 5*$mstellar$\_$err$ to ensure reliable stellar mass estimates;
iv) flux\_err > 0 and flux > 5*flux\_err (flux\_err as the corresponding uncertainty) for all specified emission lines to attain robust emission 
line features; 
v) veldisp > 5*veldisp\_err (veldisp as the velocity dispersion of the stellar component and veldisp\_err as the 
corresponding uncertainty) and veldisp between 80 and 400 km/s to ensure reliable velocity dispersion. 

Here, we explain the reason why we select galaxies with velocity dispersion between 80 and 400 km/s.
Based on SDSS documentation \footnote{
\href{https://www.sdss3.org/dr8/algorithms/veldisp.php}{https://www.sdss3.org/dr8/algorithms/veldisp.php}},
the instrumental resolution of the SDSS spectrograph is approximately 70 km/s \citep{2005ApJ...627..721G}. 
The velocity dispersion measurements distributed with SDSS spectra use template spectra convolved to a maximum sigma of 420 km/s. 
Therefore, velocity dispersion values > 420 km/s are not reliable and must not be used.
We therefore apply a selection criterion of velocity dispersion between 80 km/s and 400 km/s to obtain reliable values.
Meanwhile, since the velocity dispersion is estimated through the absorption lines included by the host galaxy features, 
galaxies with reliable velocity dispersions indicate reliable host galaxy features.

\subsection{why we assume galaxy separation to be 3$\sim$10 times $R_{NRLs}$}\label{galaxy_separation}
As demonstrated in \cite{2009ApJ...693.1554F, 2010MNRAS.407.1514E}, galaxy merging typically occurs at projected separations under 30 kpc. 
In our simulations, $R_{NLRs}$ -- estimated using the total stellar mass and the two correlations from Section 2.2 in the manuscript -- ranges from 
1.04 kpc to 3.53 kpc. 
Assuming galaxy separations of 3$\sim$10 times $R_{NLRs}$ results in this separation varying between 3.12 kpc and 35.30 kpc, similar to 
the expected value from the literature.

\subsection{why we impose a 100 km/s threshold to obtain resolvable velocity separations}\label{velocity_separation}
As demonstrated in \cite{2005ApJ...627..721G}, the SDSS spectral resolution corresponds to a resolvable velocity separation of 
$\sim$70 km/s, indicating that a velocity separation larger than 70 km/s can be detected in the SDSS spectroscopic observations. 
Therefore, we choose 100 km/s to be the minimum resolvable velocity separation in our simulations to obtain more specific results. 

\subsection{effects of considering the uncertainties of the two correlations}\label{correlation_uncertainty}
\begin{figure*}
 \includegraphics[width=0.85\paperwidth]{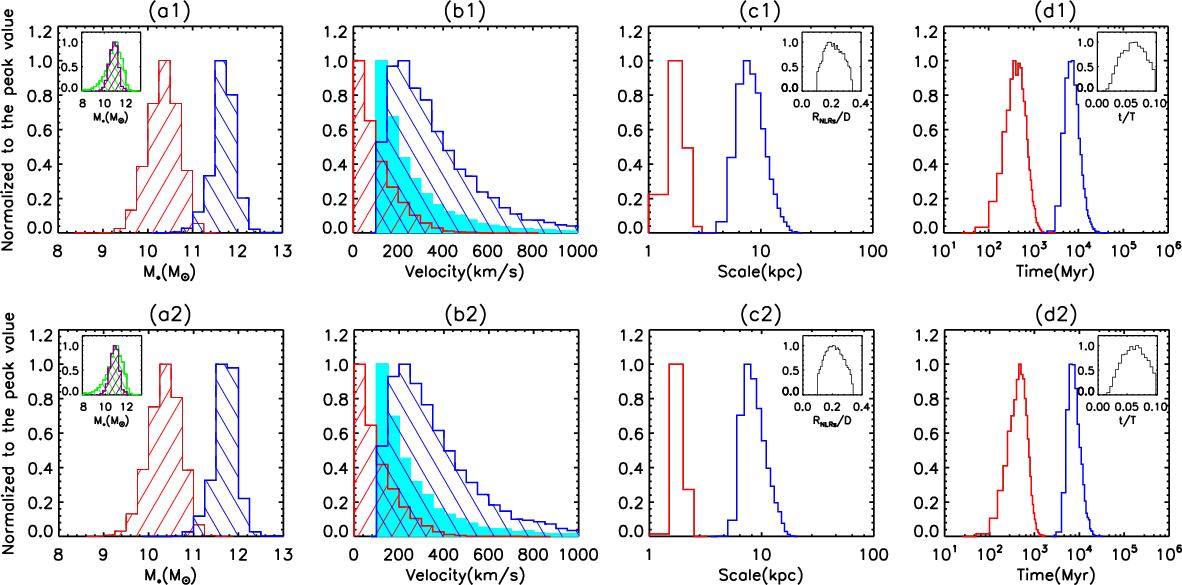}
 \centering
 \caption{
Same as Figure 2, but for simulation results with and without considering the uncertainties of the two correlations. 
The panels in the first row are for the simulation results with considering the uncertainties.
The panels in the second row are for the simulation results without considering the uncertainties.
}
 \label{simulation_uncertainty}
\end{figure*}

To explore the effects of adopting the uncertainties of the two correlations on the final results, we performed simulations with and 
without considering the uncertainties. The simulation results are shown in Figure \ref{simulation_uncertainty}. 
The results indicate that the parameter distributions of simulations that consider uncertainties become more dispersive than those of simulations 
that do not consider uncertainties, especially for $R_{NLRs}$, galaxy separation, acceleration duration, and orbital period. 
The mean velocity separations of the two kinds of simulations are similar, the former has a value of 321 km/s while the latter has a value
of 316 km/s. 
Therefore, considering the uncertainties of both correlations or not has few effects on the final simulation results. 
However, omitting the two correlations when determining $R_{NLRs}$ of the main galaxy would cause numerous invalid simulations.

\subsection{reality of the configuration of the gravitational acceleration model}\label{model_reality}
In fact, we must state that we cannot determine the configuration reality of the model, at least at the current stage.
Meanwhile, it is difficult to estimate the probability of detecting a merging galaxy pair consisting of an emission-line galaxy with 
two NLRs and a non-emission line galaxy. 
Defining this probability would theoretically reduce the likelihood of generating DPNELs shifted in the same direction, since the
probability proposed in the manuscript is estimated based on such a dual-core system model.

\section{Spectroscopic analysis of SDSS J0010}\label{spectroscopic_analysis}

\begin{figure*}
 \includegraphics[width=0.5\paperwidth]{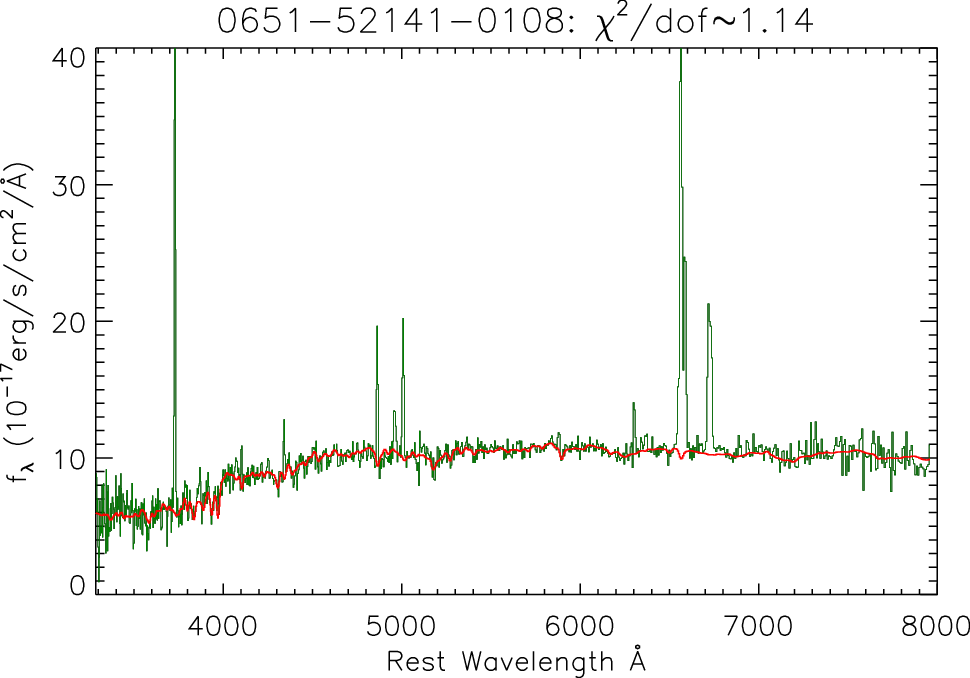}
 \centering
 \caption{The rest-frame spectrum of SDSS J0100. 
The dark green components indicate the SDSS spectrum, while the red components 
indicate the best-fitting results of the pPXF-determined host galaxy contributions. 
The mjd-plate-fiberid number of this object and the corresponding $\chi^2$/dof related to the best-fitting results are marked in the title. 
}
 \label{ppxf}
\end{figure*}

To subtract the host galaxy contributions in the spectrum, which would affect the determinations of emission line features, we adopt the penalized 
pixel fitting (pPXF) code \citep{2017MNRAS.466..798C} to determine the host galaxy features. 
The pPXF code is an improved simple stellar population (SSP) method that uses stellar population spectra while employing a 
regularization measurement. 
We have used the older version of the pPXF code to determine the host galaxy features, all emission lines are masked out during the pPXF-fitting. 
Figure \ref{ppxf} shows the SDSS provided spectrum of SDSS J0010 and the best-fitting results of pPXF-determined host galaxy 
features. 

After subtracting the host galaxy contributions, the narrow emission lines around H$\alpha$ and H$\beta$, where the Balmer emission lines, 
[N~\textsc{ii}], [S~\textsc{ii}], and [O~\textsc{iii}] are mainly considered. 
Two Gaussian functions are applied to describe each emission line, except for H$\alpha$ and [N~\textsc{ii}] doublet, which require three 
Gaussian functions for each of these emission lines due to the possible extended components covered in the spectrum. 
Through fitting the emission lines simultaneously with the \texttt{MPFIT} package, a Levenberg-Marquardt least-squares minimization 
technique, we determine the best-fitting results to the emission lines, shown in Figure \ref{spectrum}, and then obtain the parameters of 
the Gaussian components. 
Therefore, we obtain the measured velocity parameters of DPNELs of SDSS J0010 based on the central wavelengths and corresponding uncertainties of 
the core components, with blueshifted velocities of -343.32$\pm$17.27 km/s and -41.51$\pm$36.71 km/s, and a velocity separation of 
$301.80^{+26.99}_{-9.72}$ km/s. 
Here, the uncertainties are determined by the 1$\sigma$ errors of the central wavelengths related to the core components. 
It is worth noting that after carefully checking the reported DPNELs galaxies, SDSS J0010 is covered in 
\cite{2012ApJS..201...31G} and exhibits DPNELs shifted in the same direction with blueshifted velocities 
of 44.9±8.4 km/s and 316.3±2.8 km/s, and a velocity separation of 271.4±11.2 km/s in their analysis.
Our derived velocity parameters are similar to these literature values. 

\begin{figure*}
 \includegraphics[width=0.75\paperwidth]{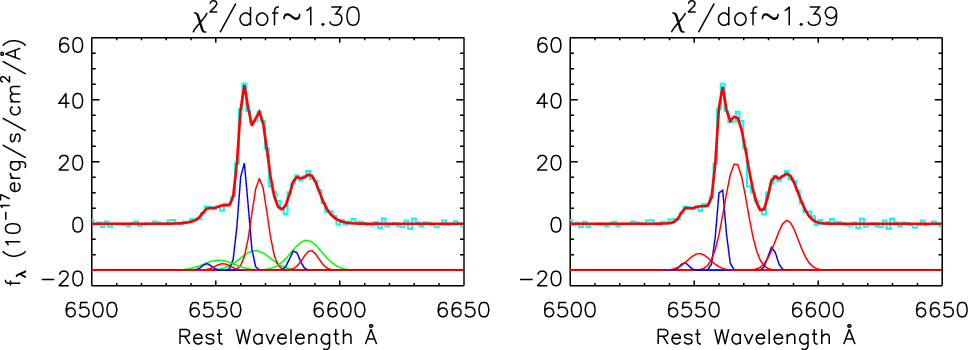}
 \centering
 \caption{
The best-fitting results to H$\alpha$ and [N~\textsc{ii}] emission lines using three Gaussian functions (model B, left panel) to fit 
each narrow emission line and two Gaussian functions (model A, right panel) to fit each narrow emission line. 
In both panels, the cyan components represent the spectrum after subtracting the host galaxy features; the thick red components represent the best-fitting results 
for emission lines; the thin red and thin blue Gaussian components from left to right indicate the double-peaked features of narrow 
[N~\textsc{ii}]$\lambda$ 6550\AA, H$\alpha$, and [N~\textsc{ii}]$\lambda$ 6585\AA. 
In the left panel, the green Gaussian components indicate the extended components of [N~\textsc{ii}]$\lambda$ 6550\AA, H$\alpha$, 
and [N~\textsc{ii}]$\lambda$ 6585\AA. 
In each panel, the corresponding $\chi^2$/dof related to the best-fitting results is marked in the title. 
}
 \label{f_test}
\end{figure*}

We also test whether the model with three Gaussian functions is preferred in fitting H$\alpha$ and [N~\textsc{ii}] doublet through 
the F-test statistical technique. 
Two models are used to fit H$\alpha$ and [N~\textsc{ii}] doublet: model A adopts two Gaussian functions to fit each narrow emission line,
and model B adopts three Gaussian functions to fit each narrow emission line. 
Through the MPFIT package, the best-fitting results of the two models for H$\alpha$ and [N~\textsc{ii}] are shown in Figure \ref{f_test}. 
Based on the different $\chi^2$/dof values for model A and model B ($\chi^2_A$ $\sim$ 108.5, $\chi^2_B$ $\sim$ 94.9, $dof_{A}$ $\sim$ 136, 
and $dof_{B}$ $\sim$ 130), the calculated $F_{p}$ value is about 
\begin{equation}
F_p = \frac{\frac{\chi^2_A - \chi^2_B}{dof_A -dof_B}}{\chi^2_B/dof_B} \sim 3.11, 
\end{equation}
where $\chi^2_A$ and $\chi^2_B$ are the sums of squared residuals, and $dof_{A}$ and $dof_{B}$ are the degrees of freedom for model A and model B, respectively. 
Based on $dof_{A}$ - $dof_{B}$ and $dof_{B}$ as the number of dofs of the F-distribution numerator and denominator, the expected value from the 
statistical F-test is about 2.94  with a 99$\%$ probability. 
Since $F_p$ = 3.11 > 2.94, it is clarified that the probability is higher than 99$\%$ to support the extended components covered in the H$\alpha$ 
and [N~\textsc{ii}] doublet. 

\begin{figure*}
 \includegraphics[width=0.8\paperwidth]{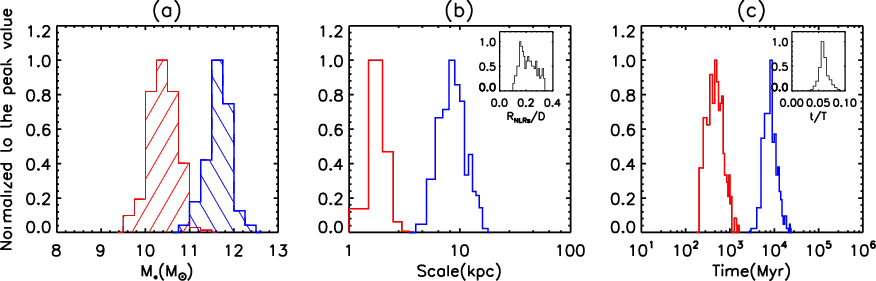}
 \centering
 \caption{The parameter distributions for the 218 satisfied runs whose velocity parameters are consistent with SDSS J0010.
Panel (a): the histograms filled in red and blue show the logarithmic total stellar mass distributions of the main and companion 
galaxies. 
Panel (b): the histograms in red and blue show the distributions of $R_{NLRs}$ and galaxy separation; the distribution 
of $R_{NLRs}$/D is shown in the top right region. 
Panel (c):  the histograms filled in red and blue show the distributions of the acceleration duration and orbital period; the distribution 
of the acceleration duration/orbital period is shown in the top right region. 
}
 \label{j0010}
\end{figure*}

Additionally, we also discuss the DPNELs shifted in the same direction of SDSS J0010 by combining our simulation results. 
Through matching the velocity parameters ($\pm$ uncertainties) of SDSS J0010 to the satisfied 10,000 simulation trials, we identified 
218 runs. The parameter distributions of these 218 simulations are shown in Figure \ref{j0010}. 
Since the parameter distributions (masses, galaxy separation, $R_{NLRs}$, acceleration duration, and orbital period) are dispersive 
rather than concentrated, it is hard to find a realistically unique set of parameters to interpret the observed DPNELs shifted in the same 
direction of SDSS J0010 in our simulation results. 

\section{Effects of different orbital trajectories}\label{different_path}

\begin{figure*}
 \includegraphics[width=0.80\paperwidth]{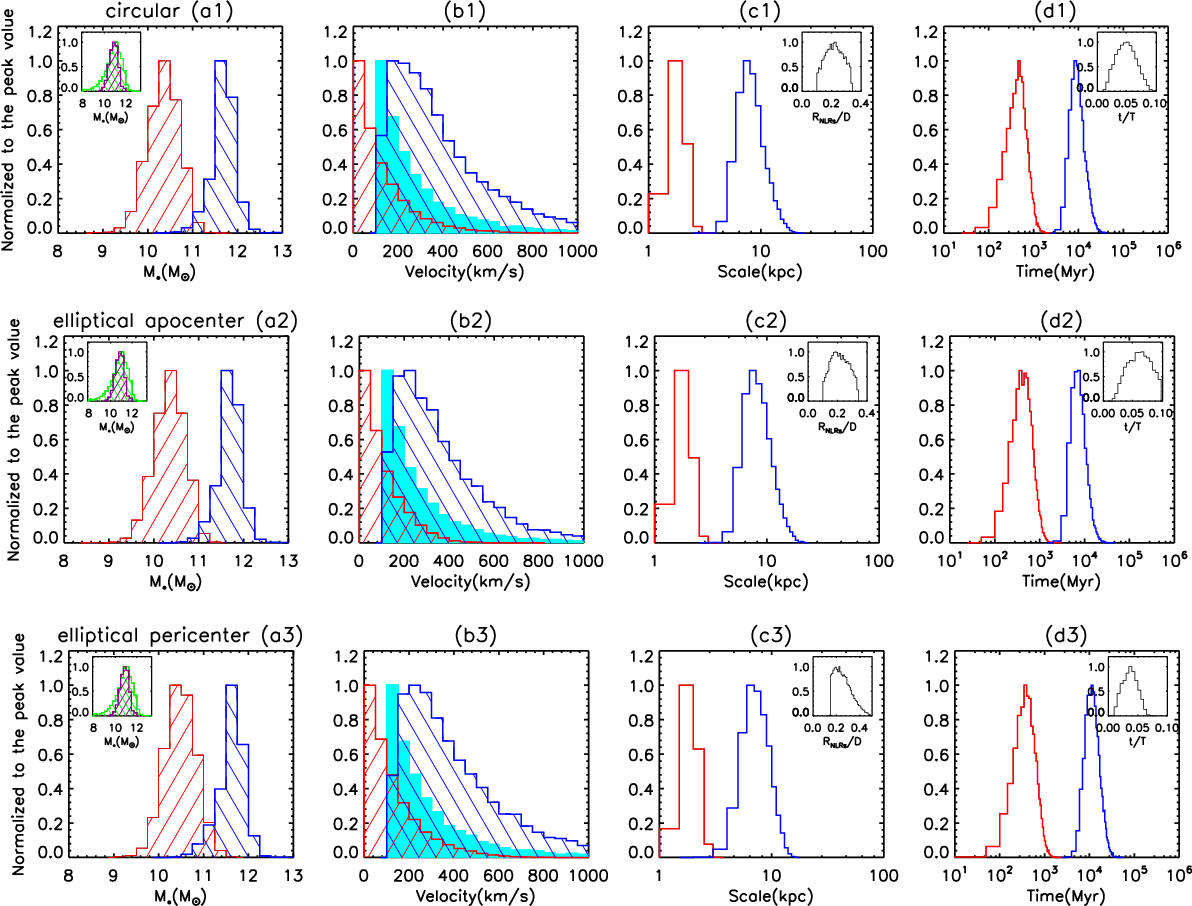}
 \centering
 \caption{Same as Figure 2. 
The panels in the first row display simulation results for a circular orbit.
The panels in the second row show simulation results for an elliptical orbit (eccentricity=0.2) with galaxies initially at apocenter. 
The panels in the third row present simulation results for an elliptical orbit (eccentricity=0.2) with galaxies initially at pericenter.
}
 \label{orbit_pic}
\end{figure*}

\begin{figure*}
 \includegraphics[width=0.5\paperwidth]{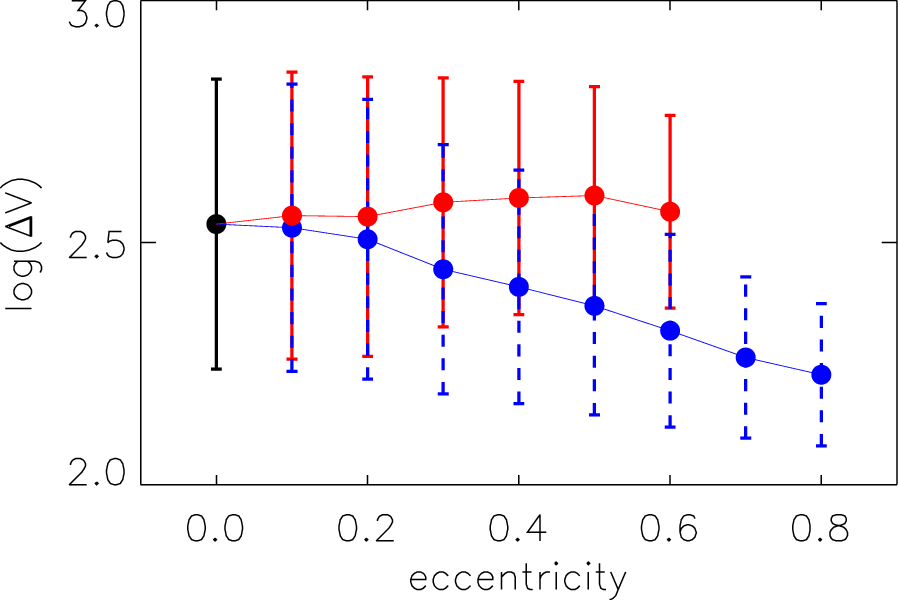}
 \centering
 \caption{On the correlation between the eccentricity of an elliptical orbit and the logarithmic mean velocity separation of DPNELs. 
The solid points in red/blue with corresponding error bars represent the results with galaxies initially at the pericenter/apocenter of the 
elliptical orbital path.
The black point represents the result with a circular orbit (eccentricity = 0). 
The error bars are calculated using the standard deviation of the corresponding velocity separations.
}
 \label{rela_e}
\end{figure*}

To explore the effects of adopting different orbital trajectories of the two galaxies in the gravitational acceleration model on the 
simulation results, we performed simulations adopting an elliptical orbital path, with the two galaxies initially positioned at both the 
apocenter and pericenter, and meanwhile in varying eccentricities.
Figure \ref{orbit_pic} presents the results for both circular orbits and elliptical orbits with an eccentricity of 0.2. 
The simulations show small differences between these two cases in the distributions of masses, $R_{NLRs}$, galaxy separation, acceleration 
duration, and orbital period. 
The relationship between eccentricity and mean peak separation of DPNELs shifted in the same direction is shown in Figure \ref{rela_e}. 
For elliptical orbits with galaxies initially at apocenter, the mean velocity separation decreases with increasing eccentricity, 
while the opposite relationship occurs for elliptical orbits with galaxies initially at pericenter.
Here, the simulations of galaxies initially at apocenter and pericenter with an identical eccentricity share the same semi-major axis. 
Therefore, since simulations with larger eccentricity result in smaller periapsis distances, the scenario with galaxies initially at pericenter 
yields no satisfied runs when eccentricity exceeds 0.6.

\end{document}